# Rational Design of High-Tc Superconductivity in the Z = 5.67 Family with Ne = 2.33

O. Paul Isikaku-Ironkwe[1,2]

[1]The Center for Superconductivity Technologies (TCST)
Department of Physics,
Michael Okpara University of Agriculture, Umudike (MOUAU),
Umuahia, Abia State, Nigeria
and
[2]RTS Technologies, San Diego, CA 92124

## Abstract

The concepts of "Rational Design of Superconductivity" from Periodic Table properties were proposed in an earlier paper (ArXiv: 1204.0233). We had shown latter too that high-Tc superconductivity exists in the Z=7.333 family with Ne=2.667, of which $MgB_2$ is a member. Here we propose and show that compounds with Z = 5.667 and Ne=2.333 will meet the conditions for high-Tc superconductivity similar to the Z = 7.333 family. The predicted Tcs for the ternary and quaternary systems of the Z =5.667 family would fall in the range 40K – 100K. We give material specific examples of some such possible rational designs of high-Tc superconductivity.

## Inspiration

"……the rational design of new superconductors will provide new directions for the grand challenge search for room-temperature superconductors, as well as materials with improved capacity to carry large currents with high critical magnetic fields".

-------- "Basic Research Needs for Superconductivity" DOE-BES (2006)

## Introduction

The discovery of superconductivity in $MgB_2$ [1] created a model of a low-Z material with high Tc. Efforts to duplicate it based only on structural and electronic symmetry failed to produce any material with identical Tc [2 - 8]. We showed earlier [9, 10] that taking into account isoelectronic and Z symmetry should lead to Tcs close to $MgB_2$'s and proposed materials[9 - 15] that should test that proposal. Using the $MgB_2$ paradigm, we investigated low-Z materials for superconductivity [16 - 20] in the Z= 4.67, 7.33, 8.67, 10.0 and 12.67 families. We have shown that though $MgB_2$ provides a model, it is not unique and that other low-Z materials with more-than-three elements do exist which we predict should have even higher Tcs. Here we propose that another family exists: the Z = 5.667 family (**F**z=5.67) with Ne=2.33. Here we show how to design it and estimate that ternary and quaternary members of this novel family of low-Z superconductors will have Tcs ranging from 40K to 100K.

## Rational Design Framework

The transition temperature Tc, of a superconductor can be estimated from certain Periodic Table parameters such as electronegativity, $\mathcal{X}$, valence electron count, Ne, atomic number, Z, and formula weight, Fw. We showed [9] that maximum Tc can be expressed as:

$$\mathbf{T_c} = \mathcal{X} \frac{Ne}{\sqrt{Z}} K_o \tag{1}$$

The value of Ko = n(Fw/Z) where n can be determined empirically. For magnesium diboride, n = 3.65. Other materials have n varying between 0 and 3.65 [9]. The synthesis of **F**z=5.67 family of materials can be achieved from other families as shown in equations 2 and 3 and depicted in figures 1 and 2.

$$3\mathbf{F}_{Z=4.67} + \mathbf{F}_{Z=8.67} = \mathbf{F}_{Z=5.67} \tag{2}$$

$$\mathbf{F}_{Z=4.67} + \mathbf{F}_{Z=6.67} = \mathbf{F}_{Z=5.67} \tag{3}$$

The average values of $\mathcal{X}$, Ne, Z, Ne/$\sqrt{Z}$ are computed for a material $A_pB_qC_rD_sE_t$ as:

$$\mathcal{X} = \frac{p\mathcal{X}_A + q\mathcal{X}_B + r\mathcal{X}_C + s\mathcal{X}_D + t\mathcal{X}_E}{p+q+r+s+t} \tag{4}$$

$$Ne = \frac{pNe_A + qNe_B + rNe_C + sNe_D + tNe_E}{p+q+r+s+t} \tag{5}$$

$$Z = \frac{pZ_A + qZ_B + rZ_C + sZ_D + tZ_E}{p+q+r+s+t} \tag{6}$$

The formula weight, Fw, of $A_pB_qC_rD_sE_t$ is expressed as:

$$Fw = pFw_A + qFw_B + rFw_C + sFw_D + tFw_E \tag{7}$$

The average values of electronegativity, $\mathcal{X}$, valence electron count, Ne, and atomic number, Z, in equations (4), (5) and (6), and the formula weight (Fw) in equation (7) are used to compute the material specific characteristics dataset (MSCD) of the materials presented here. The new material's MSCD [9] will be:

$$\text{MSCD of Material} = \langle x, Ne, Z, Ne/\sqrt{Z}, Fw, Fw/Z \rangle \tag{8}$$

$$= \langle x, 2.333, 5.667, 0.9801, Fw, Fw/Z \rangle \tag{9}$$

Here one of the materials has Ne = 1.333 and the others Ne=2.667.

In reference [9] we showed that a high-Tc should be expected when:

$$0.8 < Ne/\sqrt{Z} < 1.0 \tag{10}$$

## Synthesis Results & Tc Estimation

Tables 1, 2 and 3 show the MSCDs respectively of $F_{z=4.67}$, $F_{z=8.67}$ and $F_{z=6.67}$ families of materials. Combinatorial synthesis of a $F_{z=8.67}$ material with Ne=1.333 and three $F_{z=4.67}$ materials with Ne=2.667, lead to the Fz=5.667 family with Ne=2.333 as shown in figure 1. For example $CaLi_2$ and three moles of LiBC may combine under appropriate conditions to yield $Li_5CaB_3C_3$. The MSCD of $CaLi_2$ is given in Table 2. The MSCD of $Li_5CaB_3C_3$, from equations (8) and (9) is given as:

MSCD of $Li_5CaB_3C_3$ = ⟨1.625, 2.333, 5.667, 0.9801, 143.24, 25.28⟩ (11)

With $Ne/\sqrt{Z} = 0.9801$ we expect that the formed material will be a superconductor according to (10) above. The Tc can be estimated from equation (1) and is:

Tc = (1.625 x 0.9801 x 25.28)n = 40.3n (12)

For n= (1, 1.5, 2, 2.5, 3), Tc takes values of 40.3K, 60.4K, 80.5K, 100.6K, 120.8K respectively. In reference 9, we established that there is a strong correlation between Tc and Fw/Z with Tc increasing as FW/Z increases. From figure 2 graph of reference [9], which shows a plot of Tc vs Fw/Z, a Fw/Z of 25.3 gave a maximum Tc of about 101.1K. Thus n =2.5 for this material and the Tc is around 101K. Table 4 shows the MSCDs of some of the combinatorial synthesis possibilities and their estimated Tcs. Similarly, $F_{z=5.667}$ may be achieved by the combinatorial synthesis of $F_{z=4.67}$ and $F_{z=6.67}$ as shown in figure 2 and with results shown in Table 4. The average Fw/Z by this route is 12.84 leading to an estimated Tc of 52K for n=2.5 in equation (1).

## Discussion

We have assumed that the thermodynamics of the materials will permit the formation of the end materials shown above and in Tables 4 and 5. The results herein therefore await experimental verification. If correct, they are also a verification of the assumptions on which we estimated Tc of superconductors in terms of material specific parameters of electronegativity, valence electron count, atomic number and formula weight. We invite serious experimentalists to test out this and previous predictions [10 -20] of low-Z high Tc superconductors.

## Conclusion

We showed how to design the $F_z$=5.67 family of materials with $N_e$=2.33. We found that members may have more than three elements, with Fw/Z between 12.8 and 25.3. Tcs are estimated between 40K and 100K. They are worth further experimental investigation.

## Acknowledgements

The author acknowledges financial support from M. J. Schaffer, formerly at General Atomics San Diego. Discussions with A.O.E. Animalu at University of Nigeria, M. B. Maple at UC San Diego and J. R. O'Brien at Quantum Design proved useful in the development of the ideas herein.

**TABLES**

| | Material | $x$ | Ne | Z | Ne/$\sqrt{Z}$ | Fw | Fw/Z |
|---|---|---|---|---|---|---|---|
| 1 | LiBC | 1.8333 | 2.6667 | 4.6667 | 1.2344 | 29.76 | 6.377 |
| 2 | BeB$_2$ | 1.8333 | 2.6667 | 4.6667 | 1.2344 | 30.63 | 6.564 |
| 3 | Be$_2$C | 1.8333 | 2.6667 | 4.6667 | 1.2344 | 30.03 | 6.435 |
| 4 | Li$_2$O | 1.8333 | 2.6667 | 4.6667 | 1.2344 | 29.88 | 6.403 |
| 5 | LiBeN | 1.8333 | 2.6667 | 4.6667 | 1.2344 | 29.96 | 6.420 |

**Table 1:** 5 MSCDs of Z =4.67 materials with Ne=2.67

| | Material | $x$ | Ne | Z | Ne/$\sqrt{Z}$ | Fw | Fw/Z |
|---|---|---|---|---|---|---|---|
| 1 | Li$_2$Ca | 1.0 | 1.333 | 8.667 | 0.0.4529 | 53.96 | 6.23 |

**Table 2:** MSCD of Li$_2$Ca, a Z =8.67 material with Ne=1.333

| | Material | $x$ | Ne | Z | Ne/$\sqrt{Z}$ | Fw | Fw/Z |
|---|---|---|---|---|---|---|---|
| 1 | NaBeB | 1.467 | 2.0 | 6.667 | 0.7746 | 42.81 | 6.42 |

**Table 3:** MSCD of NaBeB, a Z =6.67 material with Ne=2.0. Combining it with Z=4.67 materials in Table 1 will give the results in Table 4.

| | Material | $x$ | Ne | Z | Ne/$\sqrt{Z}$ | Fw | Fw/Z |
|---|---|---|---|---|---|---|---|
| 1 | LiNaBeB$_2$C | 1.65 | 2.333 | 5.667 | 0.9801 | 72.57 | 12.81 |
| 2 | Li$_2$NaBeBO | 1.65 | 2.333 | 5.667 | 0.9801 | 72.69 | 12.83 |
| 3 | Na Be$_3$BC | 1.65 | 2.333 | 5.667 | 0.9801 | 72.84 | 12.85 |
| 4 | NaBe$_2$B$_3$ | 1.65 | 2.333 | 5.667 | 0.9801 | 73.44 | 12.96 |
| 5 | LiNaBe$_2$BN | 1.65 | 2.333 | 5.667 | 0.9801 | 72.77 | 12.84 |

**Table 4:** 5 MSCDs of Z =5.67 materials with Ne=2.33 and Ne/$\sqrt{Z}$ =0.9801. With an average Fw/Z=12.84, Tc is estimated in the range 21K to 52K, using equation (1).

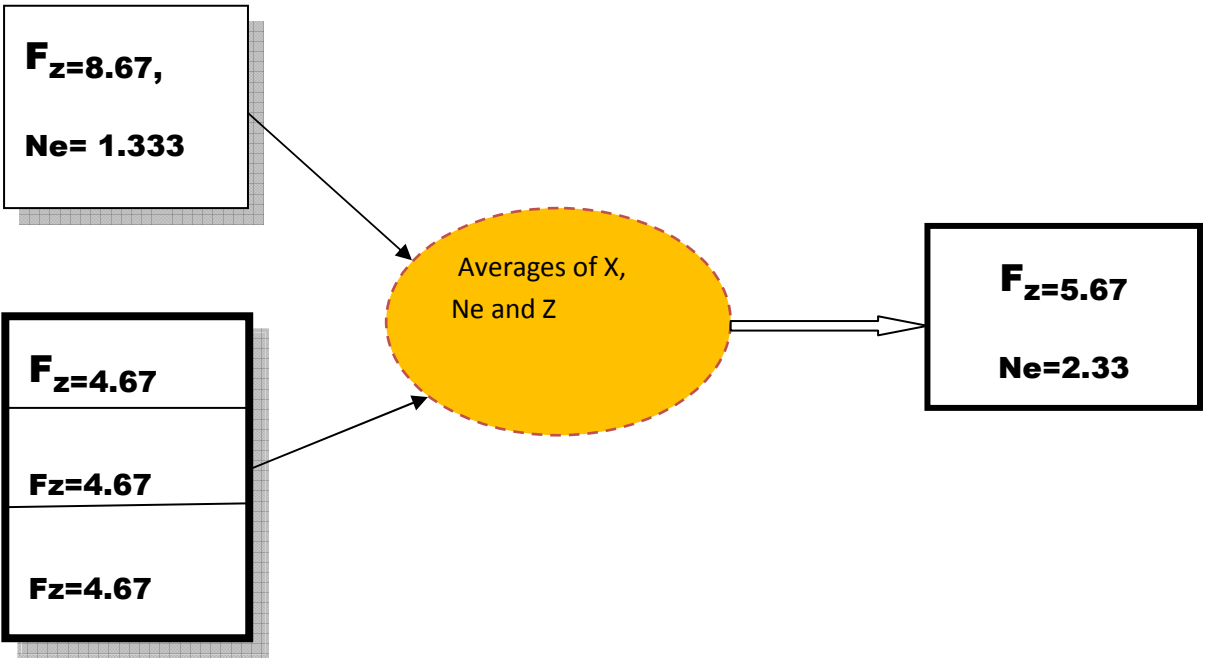

**Figure 1:** Combinatorial synthesis of Fz=5.67 family of materials from Fz=8.67 and THREE Fz=4.67 materials. One of the materials must have Ne=1.333. We create 3-,4- and 5- elements 6-atom systems with Ne=2.33 and Fw/Z of around 25.3.

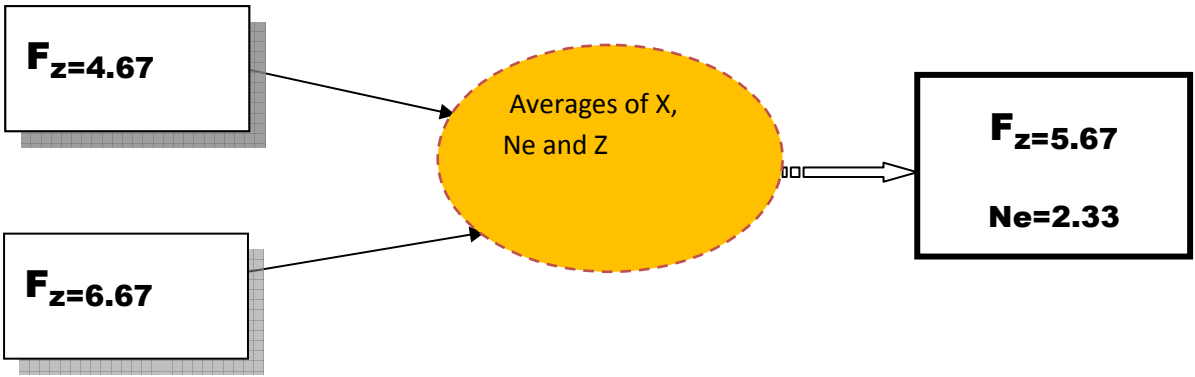

**Figure 2:** Combinatorial synthesis of Fz=5.67 family of materials from Fz=4.67 and Fz=6.67. We create 3-, 4- and 5-elements 6-atom systems with Ne=2.33 and Fw/Z of around 12.8.